\documentclass[twocolumn,aps,floatfix]{revtex4}
\usepackage{amssymb}
\usepackage{graphicx}
\usepackage{amsmath}
\usepackage{pstricks}

\newcommand{\B}{\hat{a}_0}
\newcommand{\Bd}{\hat{a}_0^\dagger}
\newcommand{\C}{\hat{a}_1}
\newcommand{\Cd}{\hat{a}_1^\dagger}
\newcommand{\D}{\hat{a}_2}
\newcommand{\Dd}{\hat{a}_2^\dagger}

\begin{document}
%\draft
\title{Tunable dipolar resonances and Einstein-de Haas effect in a $^{87}$Rb atoms condensate}
\author{\mbox{Tomasz \'Swis{\l}ocki$^{1}$, Tomasz Sowi\'nski$^{1}$, Joanna Pietraszewicz$^{1}$, Miros{\l}aw Brewczyk$^2$},  Maciej Lewenstein$^{3,4}$, Jakub Zakrzewski$^{5,6}$, and Mariusz Gajda$^{1}$}
\affiliation{
\mbox{$^1$Instytut Fizyki PAN, Al. Lotnik\'ow 32/46, 02-668 Warszawa, Poland}\\
\mbox{$^2$Wydzia{\l} Fizyki, Uniwersytet w Bia\l ymstoku, ul. Lipowa 41, 15-424 Bia\l ystok, Poland} \\
\mbox{$^3$ICFO - Institut de Ci\`ences Fot\`oniques, Parc Mediterrani de la Tecnologia, E-08860
Castelldefels, Barcelona, Spain} \\
\mbox{$^4$ ICREA - Instituci\'o Catalana de Recerca i Estudis Avan\c cats, 08010 Barcelona, Spain}
\mbox{$^6$Instytut Fizyki imienia Mariana Smoluchowskiego, Uniwersytet Jagiellonski, ulica Reymonta 4, 30-059 Krak\'ow, Poland}\\
\mbox{$^7$Mark Kac Complex Systems Research Center, Jagiellonian University, ulica Reymonta 4, 30-059 Krak\'ow, Poland}
}
\date{\today}

\begin{abstract}
We study a spinor condensate of $^{87}$Rb atoms in $F=1$ hyperfine state confined in an optical dipole trap. Putting initially all atoms in $m_F=1$ component we observe a significant transfer of atoms to other, initially empty  Zeeman states exclusively due to dipolar forces. Because of conservation of a total angular momentum the atoms going to other Zeeman components acquire an orbital angular momentum and circulate around the center of the trap. This is a realization of Einstein-de Haas effect in a system of cold gases. We show that the transfer of atoms via dipolar interaction is possible only when the energies of the initial and the final sates are equal. This condition can be fulfilled utilizing a resonant external magnetic field, which tunes energies of involved states via the linear Zeeman effect. We found that there are many final states of different spatial density which can be tuned selectively to the initial state. We show a simple model explaining high selectivity and controllability of weak dipolar interactions in the condensate of $^{87}$Rb atoms.

\end{abstract}

\maketitle

\section{Introduction}
\label{introduction}

Experimental achievement of Bose-Einstein condensation in a gas of $^{52}$Cr atoms \cite{52Cr1,52Cr2} has launched a huge interest in studying properties of ultracold dipolar systems \cite{Lewenstein_review,Ueda_review}. Although polar molecules with their large electric dipole moments seem to be perfect candidates to investigate dipolar effects such systems have not been condensed so far. Also the necessity of an external field inducing the dipole in these systems might suppress some intrinsic interaction effects (for instance, the electric dipole-dipole interactions do not conserve the total angular momentum, i.e., the sum of spin and orbital angular momentum). On the other hand, experiments with chromium condensates already showed spectacular features characteristic for dipolar interactions. For example, in Ref. \cite{Pfau_expansion} the suppression (and even complete inhibition) of inversion of ellipticity during expansion of a cloud of chromium atoms was observed. In Ref. \cite{Pfau_collapse}, on the other hand, the collapse dynamics of $^{52}$Cr condensate was studied. The scattering length was decreased by means of a Feshbach resonance below the critical value above which the condensate is stable. Then the system was hold in a trap for some time and eventually the trap was switched off. The specific cloverleaf patterns for the density caused by anisotropic dipole-dipole interactions were observed when the system was imaged after time of flight.

Some authors suggested, however, that the magnetic dipolar interactions might already lead to observable effects in condensates of alkali atoms \cite{Pu, Ueda_1,KG,You,Yi,Saito,TS}, whose magnetic dipole moment is an order of magnitude lower than that of chromium atoms. Indeed, in experimental work of Ref. \cite{Kurn} it is demonstrated that a spin-$1$ $^{87}$Rb spinor condensate exhibits dipolar properties. The authors of \cite{Kurn} observe the spontaneous decay of helical spin textures toward a spatially modulated structure of spin domains and relate this effect to the presence of magnetic dipolar interactions in the system. This claim is supported by an observation that the modulated phase is suppressed when the dipolar interactions are reduced. There are, however, some difficulties in understanding the results of this experiment within the frame of mean-field and Bogoliubov theories at zero temperature \cite{Kawaguchi}.

One of the most spectacular manifestations of dipolar interactions is the Einstein-de Haas effect \cite{EdH} studied recently theoretically in gaseous systems of chromium \cite{Ueda_2,Pfau} and rubidium condensates \cite{KG}. Interestingly, this phenomenon might occur even in very weak dipolar systems. To observe the Einstein-de Haas effect in alkali atoms condensates one has to tune an external magnetic field to some resonant value or for a fixed magnetic field to modify appropriately trapping parameters. Only when the Zeeman energy fits the energy gap between different single-particle spatial states associated with $m_F=1$ and $m_F=0$ spin states, the transfer of atoms is resonantly amplified.

We note that the resonant magnetic field is typically of the order of tens or hundreds micro-Gauss making the observation of the Einstein-de Haas effect difficult at present. Since dipole-dipole interactions are very weak the resonant curves are also very narrow. This means that experimental realization needs high precision. On the other hand, it guarantees that resonances are highly selective. By choosing an appropriate value of the magnetic field one can tune the transition of atoms to particular spatial state. Indeed, controlling dipolar interactions is the crucial point in working with dipolar systems. Such a control has been recently imposed in chromium condensate \cite{French1}. It was shown that the external static magnetic field strongly influences the dipolar relaxation rate -- there exists a range of magnetic field intensities where this relaxation rate is strongly reduced allowing for the accurate determination of $S=6$ scattering length for chromium atom. In Ref. \cite{French2}, on the other hand, two-dimensional optical lattice and a static magnetic field are used to control the dipolar relaxation into higher lattice bands. In this work an evidence for the existence of the relaxation threshold with respect to the intensity of the magnetic field is given. As the authors of Ref. \cite{French2} claim such an experimental setup might lead to the observation of the Einstein-de Haas effect.

This paper is organized as follows. In Sec. \ref{description} we describe the system under consideration. In particular, we discuss the contact and dipolar interactions (we refer the reader to Appendix \ref{first} for further details). Sec. \ref{EdH} presents numerical results regarding the resonances and the Einstein-de Haas effect in a rubidium condensate whereas Sec. \ref{explanation} introduces a simple model explaining the origin of the resonances as well as properties of spatial states populated via dipolar interactions. We end with conclusions in Sec. \ref{conclusions}. Moreover, the Appendix \ref{second} offers a lot of technical details implemented in numerical procedure used to obtain the results presented in this paper as well as in our previous work \cite{KG}.

\section{Description of the system}
\label{description}
We investigate a spinor condensate of atoms in the $F=1$ hyperfine state. In addition to the dominating binary contact interactions, we consider long-range dipolar magnetic interactions. In the formalism of second quantization, the Hamiltonian of the system is given by
\begin{eqnarray}
&&H = \int d^3r \left[ \hat{\psi}^{\dagger}_i(\boldsymbol{r}) \left(-\frac{\hbar^2}{
2 m}
\nabla^2 + V_{ext}(\boldsymbol{r}) \right)   \hat{\psi}_i(\boldsymbol{r})
\right.  \nonumber  \\
&&\left. -\gamma \hat{\psi}^{\dagger}_i(\boldsymbol{r})\, \boldsymbol{B}\!\cdot\!\boldsymbol{F}_{ij}\,
\hat{\psi}_j(\boldsymbol{r}) + \frac{c_0}{2}\, \hat{\psi}^{\dagger}_j(\boldsymbol{r})
\hat{\psi}^{\dagger}_i(\boldsymbol{r})
\hat{\psi}_i(\boldsymbol{r}) \hat{\psi}_j(\boldsymbol{r}) \right.   \nonumber  \\
&&\left. +\frac{c_2}{2}\, \hat{\psi}^{\dagger}_k(\boldsymbol{r})
\hat{\psi}^{\dagger}_i(\boldsymbol{r})\,
\boldsymbol{F}_{ij}\!\cdot\!\boldsymbol{F}_{kl}\, \hat{\psi}_j(\boldsymbol{r}) \hat{\psi}_l(\boldsymbol{
r})
\right]   \nonumber  \\
&&+ \frac{1}{2}\int d^3r\, d^3r' \hat{\psi}^{\dagger}_k(\boldsymbol{r})
\hat{\psi}^{\dagger}_i(\boldsymbol{r}') V^d_{ij,kl}(\boldsymbol{r}-\boldsymbol{r}')
\hat{\psi}_j(\boldsymbol{r}') \hat{\psi}_l(\boldsymbol{r})   \,,  \nonumber  \\
\label{Ham}
\end{eqnarray}
with repeated indices being summed over the values $+1$, $0$, and $-1$. The field operator $\hat{\psi}_i(\boldsymbol{r})$ ($\hat{\psi}_i^{\dagger}(\boldsymbol{r})$) annihilates (creates) an atom in the hyperfine state $|F=1,i\rangle$ at point $\boldsymbol{r}$. The first row in (\ref{Ham}) represents the single-particle Hamiltonian that consists of the kinetic energy part and the trapping potential assumed to be independent of the hyperfine state. The first term in the second row describes the interaction with the magnetic field $\boldsymbol{B}$ which is in our case always directed along the $z$-axis. Magnetic moment of the atom $\boldsymbol{\mu}$ if proportional to the dimensionless spin-1 matrices $\boldsymbol{F}$, $\boldsymbol{\mu}=\mu \boldsymbol{F}$. Effective magnetic moment of the atom $\mu$ differs from the Bohr magneton by Lande giromagnetic factor $\gamma$, $\mu=\gamma \mu_B$. The terms with coefficients $c_0$ and $c_2$ describe the spin-independent and spin-dependent parts of the contact interactions, respectively. Coefficients $c_0$ and $c_2$ can be expressed with the help of the scattering lengths $a_0$ and $a_2$ which determine the collision of atoms in a channel of total spin $0$ and $2$. One has $c_0=4\pi\hbar^2(a_0+2a_2)/3m$ and $c_2=4\pi\hbar^2(a_2-a_0)/3m$ ($m$ is the mass of an atom) \cite{Ho}, where $a_0=5.387$\,nm and $a_2=5.313$\,nm \cite{a0a2}. Finally, the last term represents the magnetic dipolar interactions and originates from the interaction energy of two magnetic dipole moments $\boldsymbol{\mu}_1$ and $\boldsymbol{\mu}_2$
\begin{eqnarray}
V^d=\frac{\boldsymbol{\mu}_1\! \cdot\! \boldsymbol{\mu}_2}{|\boldsymbol{r}-\boldsymbol{r}'|^3}-3
\frac{
[\boldsymbol{\mu}_1\! \cdot\! (\boldsymbol{r}-\boldsymbol{r}')] \,
[\boldsymbol{\mu}_2\! \cdot\! (\boldsymbol{r}-\boldsymbol{r}')]}
{|\boldsymbol{r}-\boldsymbol{r}'|^5}   \;,
\label{mm}
\end{eqnarray}
where $\boldsymbol{r}$ and $\boldsymbol{r}'$ denote the positions of the dipoles.

The equation of motion reads
\begin{eqnarray}
i\hbar \frac{\partial}{\partial t}
\left(
\begin{array}{l}
\hat{\psi}_1 \\
\hat{\psi}_0 \\
\hat{\psi}_{-1}
\end{array}
\right)
=
({\cal{H}}^{sp} + {\cal{H}}^c + {\cal{H}}^d )
\left(
\begin{array}{l}
\hat{\psi}_1 \\
\hat{\psi}_0 \\
\hat{\psi}_{-1}
\end{array}
\right)   \,\,,
\label{Eqmot}
\end{eqnarray}
where ${\cal{H}}^{sp}=-\frac{\hbar^2}{2 m} \nabla^2 + V_{ext} - \mu m_F B$ represents the single-particle part (including the interaction with the external magnetic field) whereas ${\cal{H}}^cc$ and ${\cal{H}}^d$ originate from the two-particle interactions and are discussed in detail in the Appendix \ref{first}. Here we just state that the diagonal part of ${\cal{H}}^c$ is responsible for elastic collisions whereas other elements of ${\cal{H}}^c$ allow for the change of spin projections of individual atoms preserving, however, the projection of total spin.

The dipolar part of Eq. (\ref{Eqmot}), ${\cal{H}}^d$, is more complex. To better understand the dipolar processes it is convenient to rewrite the dipolar interactions in the following form \cite{PethickSmith}
\begin{equation}
V^d \propto  %\frac{1}{r^3}
\sum_{\mu=-2}^{\mu=2}Y_{2\mu}^{\star}(\boldsymbol{\hat{r}})\Sigma_{2\mu}   \;,
\label{expansion}
\end{equation}
where $Y_{2\mu}(\boldsymbol{\hat{r}})$ (with $\boldsymbol{\hat{r}}$ denoting a unit vector in the direction of relative position of two atoms) is a spherical harmonics of rank-2 and $\Sigma_{2\mu}$ defined as
\begin{eqnarray}
\Sigma_{2,0} =&& -\sqrt{\frac{3}{2}}(F_{1z}F_{2z}-\boldsymbol{F}_1 \cdot\boldsymbol{F}_2 /3)
\nonumber \\
\Sigma_{2,\pm1} =&& \pm\frac{1}{2}(F_{1z}F_{2\pm}+F_{1\pm}F_{2z})  \nonumber\\
\Sigma_{2,\pm2} =&& -\frac{1}{2}F_{1\pm}F_{2\pm}
\label{tensor}
\end{eqnarray}
is a rank-2 spherical tensor built of atomic spin operators (including raising and lowering operators $F_{1\pm}$ and $F_{2\pm}$). It is clear from (\ref{expansion}) that when two atoms interact the total spin     projection $M_F$ (as well as the total spin itself) can change at most by $2$ whereas the spin projection of individual atoms (see (\ref{tensor})) changes maximally by $1$ not by $2$. Therefore, the atom can not be transferred directly from the $m_F=1$ to $m_F=-1$ component (populating the $m_F=-1$ state is a second order process). When $\Delta M_F=\pm2$, the last row in (\ref{tensor}) shows that the projection of the spin of each atom changes by $+\hbar$ or $-\hbar$, i.e., both atoms initially in the same state go simultaneously to the nearest (in a sense of magnetic number $m_F$) state or when atoms are in different but neighboring components they are transferred to the states shifted in number $m_F$ by $+1$ or $-1$. In addition to the processes just described there are the atomic collisions in which only one atom is transferred to the other Zeeman state ($\Delta M_F=\pm1$, see the middle row in (\ref{tensor})) or the collisions that do not change the spin projection ($\Delta M_F=0$, the first row in (\ref{tensor})). In the latter case, however, still the process transferring one atom from $m_F=0$ to $m_F=1$ and the other from $m_F=0$ to $m_F=-1$ states is allowed just like it is allowed for contact interactions. According to (\ref{expansion}) in each case the change of the total spin projection is accompanied by an appropriate change of a relative orbital angular momentum of colliding atoms.

Hence, the dipolar interactions do not conserve the projection of total spin of two interacting atoms. Neither the projection of total orbital angular momentum is preserved (see (\ref{com})). However, the dipolar interactions couple the spin and the orbital motion of atoms (with regard to any axis) as revealed by the last relation in (\ref{com})
\begin{eqnarray}
 &&[V^d,F_{1z}+F_{2z}] \neq 0,\;\;\;\;   [V^d,L_{1z}+L_{2z}] \neq 0,
 \nonumber  \\
 &&[V^d,L_{1z}+L_{2z}+F_{1z}+F_{2z}]=0  \,.
 \label{com}
\end{eqnarray}
Therefore, going to $m_F=0$ and $m_F=-1$ states atoms acquire the orbital angular momentum and start to circulate around the center of the trap. This is the essence of the famous Einstein-de Haas effect and the discussion in Sec. \ref{EdH} suggests its possible realization in cold gases.

\section{Dipolar resonances and the Einstein-de Haas effect}
\label{EdH}

In this Section we investigate the properties of dipolar spinor condensate of $^{87}$Rb
atoms in the $F=1$ hyperfine state ($\gamma = 1/2$) by solving the mean-field version of Eq. (\ref{Eqmot})
\begin{eqnarray}
i\hbar \frac{\partial}{\partial t}
\left(
\begin{array}{l}
\psi_1 \\
\psi_0 \\
\psi_{-1}
\end{array}
\right)
=
({\cal{H}}^{sp}+ {\cal{H}}^c + {\cal{H}}^d )
\left(
\begin{array}{l}
\psi_1 \\
\psi_0 \\
\psi_{-1}
\end{array}
\right)   \;
\label{meanfield}
\end{eqnarray}
with $\psi_i$ being the wave function (referred sometimes also to as an order parameter) of the $i$-th spin component.
Usually, dipolar properties of rubidium condensates are neglected. This is because the magnetic moment of rubidium atoms is small. Comparing typical energy related to the dipolar interactions (given by $\mu^2 n$, where $\mu$ is the magnetic moment of $^{87}$Rb atom and $n$ is the atomic density) to the characteristic contact interactions energy $gn$ ($g$ determines the strength of the contact interactions) one obtains for the maximally stretched $F=1$ hyperfine state (i.e., for the spin-polarized case where $g=4\pi\hbar^2 a_2/m$) of $^{87}$Rb the ratio of the order of $10^{-4}$. This is two orders of magnitude lower than the corresponding ratio calculated for condensate of chromium atoms which is commonly considered as a dipolar condensate. In this Section we show, however, that for a rubidium condensate there exist resonances which significantly enhance the effect of dipolar interactions and might lead to the manifestation of various properties in as strong way as in the case of chromium condensate.

The numerical experiment, we perform, consists in preparing the condensate initially in $m_F=1$ Zeeman state (all atomic magnetic moments aligned along the magnetic field). It is done by evolving the mean-field version of Eq. (\ref{Eqmot}) in imaginary time at the presence of external magnetic which is large enough to enforce almost all atoms to populate $m_F=1$ component. Next, we change both the direction and the value of the magnetic field. What we usually observe is the transfer of small number of atoms to $m_F=0$ and $m_F=-1$ components. We would like to stress that in our numerical experiment the initial condition excludes the short-range spin dynamics and the depletion of $m_F=1$ state occurs only due to dipolar interactions.

\begin{figure}[!thb]
%\resizebox{3.5in}{2.2in} {\includegraphics{normy23_50k_ART.eps}}
\resizebox{3.5in}{2.2in} {\includegraphics{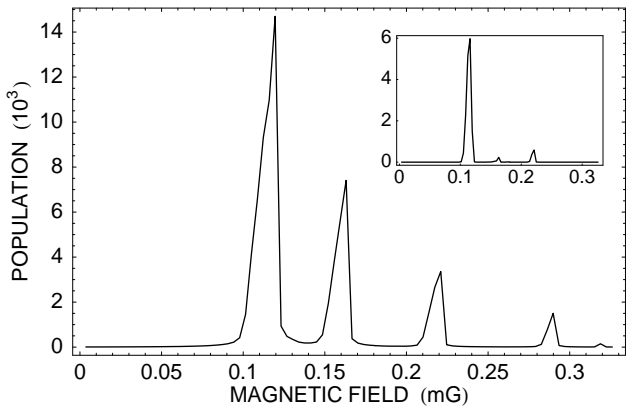}}
\caption[]{Population of $m_F=0$ (main frame) and $m_F=-1$ (inset) components as a function of magnetic field (note that the magnetic field is pointed towards negative $z$-axis and only its values are depicted here). We observe many magnetic resonances corresponding to different values of magnetic field. Initial number of atoms in $m_F=1$ state is equal to $N=5 \times 10^4$ and the aspect ratio is $\beta=\omega_z/\omega_r=1/4$. Maximal transfer is reached at $t \sim 0.1$s for each resonance.
\label{res1}}
\end{figure}

\begin{figure}[!thb]
%\resizebox{3.2in}{2.in} {\includegraphics{normy_rez1_50k_ART.eps}}
\resizebox{3.4in}{2.2in} {\includegraphics{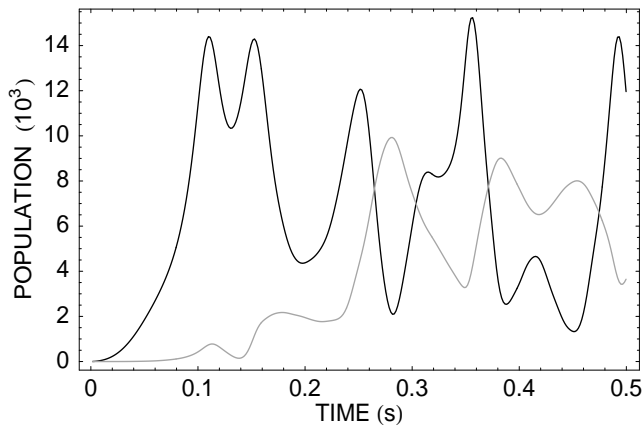}}
\caption[]{Population of $m_F=0$ (black) and $m_F=-1$ (gray) components as a function of time for the first resonance. Magnetic field is equal to $B=-0.12$mG.
\label{res12}}
\end{figure}

However, there exist the particular values of the magnetic field (for example, $B\approx -0.11$\,mG for the axially symmetric trap with $\omega_z=2\pi \times 100$\,Hz and $\omega_r=2\pi \times 400$\,Hz and initial number of atoms $N=5\times 10^4$, as it is shown in Fig. \ref{res1}) when the observed transfer is large -- of the order of $10\%$ of initial number of atoms in $m_F=1$ state. The maximal transfer happens after the time of the order of $100$\,ms (see Fig. \ref{res12}) and is consistent with the characteristic time scale related to dipolar interaction, which is $\hbar/\mu^2 n$. This formula tells us that increasing the atomic magnetic moment (for example, by going from rubidium to chromium condensate) the time needed for reaching the maximal transfer gets shorter. Indeed, for chromium condensate (with $12$ times larger the magnetic moment in comparison with $^{87}$Rb atoms in $F=1$ state) this time is of the order of $1$\,ms \cite{Ueda_2}. Similar behavior is visible for larger magnetic fields.

%fig_1 rezonanse, beta=4

%fig_2 obsadzenia_1

%fig_3 gestosci XY, beta=4
\begin{figure}[!htb]
%\resizebox{3.4in}{2.2in} {\includegraphics{dens_faz_50k_rez1_556.eps}}
\resizebox{3.4in}{2.2in} {\includegraphics{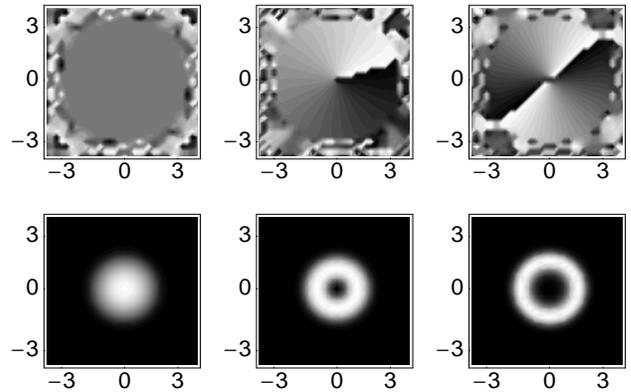}}
\caption[]{Phase (upper panel) and density (lower panel) of $m_F=+1,0,-1$ components (left, middle, and right columns, respectively) in $xy$ plane for $N_{+1} = 5 \times 10^4$ and $\beta=1/4$. The presented data correspond to the first maximum in the resonance structure (see Fig. \ref{res1}). Magnetic field is equal to $B=-0.12$mG. $N_0=15000$ and $N_{-1}=6000$ atoms both at their maximal transfers.
\label{denxy}}
\end{figure}

%fig_4 gestosci XZ, m_f=0, beta=4
\begin{figure}[!htb]
\begin{center}
%\resizebox{3.5in}{3.5in} {\includegraphics{dens_23_50k_rez1-4_XZ_556.eps}}
\resizebox{3.5in}{2.9in} {\includegraphics{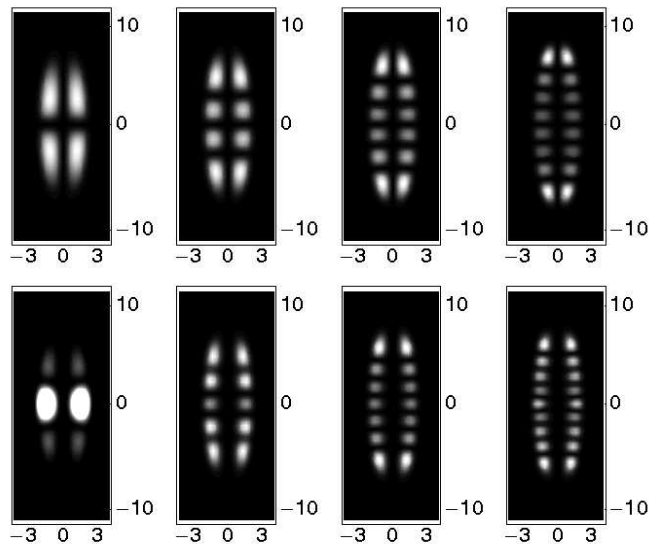}}
\caption[]{Density of $m_F=0$ (upper panel) and $m_F=-1$ (lower panel) components in $xz$ plane for $N_{+1} = 5 \times10^4$ atoms and $\beta=1/4$. Successive frames show density patterns characteristic for resonances depicted in Fig. \ref{res1}. Note the increasing number of rings with the order of the resonance as well as the presence of even (odd) number of rings in $m_F=0$ ($m_F=-1$) component.
\label{denxz0}}
\end{center}
\end{figure}

An interesting observation is made when looking at the phase and the density of spinor components for all resonances we found (Figs. \ref{denxy} and \ref{denxz0}). First of all, the inspection of the phase of the wave functions of $m_F=0,-1$ components proves the formation of quantized vortices in these components. For $m_F=0$ state the phase of the order parameter winds up by $2\pi$ (upper middle frame in Fig. \ref{denxy}) whereas in $m_F=-1$ component (upper right frame in Fig. \ref{denxy}) the phase winds up by $4\pi$. At the same time the orbital angular momentum per atom equals $\hbar$ and $2\hbar$ in $m_F=0$ and $m_F=-1$ components, respectively. The circulation induced in $m_F=0$ and $m_F=-1$ states is, in fact, the realization of the Einstein-de Haas effect in cold gases. At the same time the densities exhibit non typical patterns. The density of $m_F=0$ state consists of even number of rings when looking from a side (Fig. \ref{denxz0}, upper panel) whereas in the case!
  of $m_F=-1$ component the number of rings is odd (Fig. \ref{denxz0}, lower panel). The explanation of this property will be given in Sec. \ref{explanation}, here we just present some arguments showing that the number of rings in components is strictly related to the channel the atoms choose when they collide.
  
Since the number of rings in $m_F=0$ component is even (Fig. \ref{denxz0}, upper panel), in the simplest case its wave function fulfills the condition $\psi_0 (\boldsymbol{r}) \sim Y_{l,m}$, where the difference $l-m$ is odd. The dipole-dipole interactions are small and can be treated as a perturbation. Therefore, the wave function of $m_F=-1$ component is determined by nonzero amplitudes corresponding to processes in which only one atom changes its spin projection upon the collision (such amplitudes are given by $\langle Y_{l',m'}| Y_{21}| Y_{lm} \rangle$) and processes when both atoms change their spin projections (here amplitudes are $\langle Y_{l',m'}| Y_{22}| Y_{lm} \rangle$). In the first case the amplitudes differ from zero only when $m'=1+m$ and $l'=l-2,l,l+2$ which means that $l'-m'$ is always even. Therefore, the density of $m_F=-1$ state exhibits odd number of rings. However, in the second case $l'-m'$ is odd what implies even number of rings in $m_F=-1$ component which is not the case (see Fig. \ref{denxz0}, lower panel). Evidently, the second channel is somehow closed.

\begin{figure}[!htb]
\begin{center}
%\resizebox{3.5in}{2.2in} {\includegraphics{normy_50k_SF.eps}}
\resizebox{3.5in}{2.2in} {\includegraphics{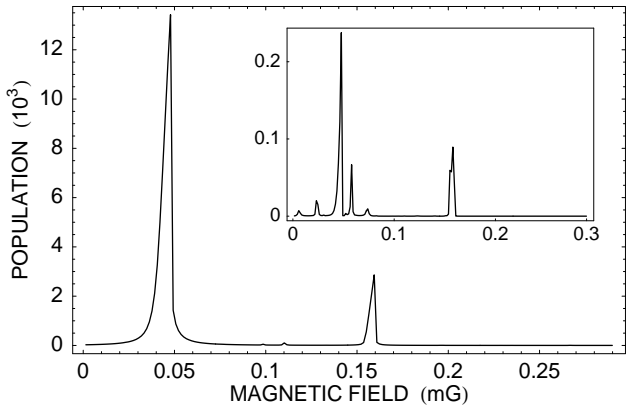}}
\caption[]{Population of $m_F=0$ (main frame) and $m_F=-1$ (inset) components as a function of magnetic field. We observe magnetic resonances corresponding to different values of magnetic field. Initial number of atoms in $m_F=1$ state is equal to $N_{+1}=5 \times 10^4$ and the aspect ratio is $\beta=1$. Maximal transfer is reached at $t \sim 0.3$s for each resonance.
\label{ss}}
\end{center}
\end{figure}

\begin{figure}[!htb]
\begin{center}
%\resizebox{3.5in}{2.3in} {\includegraphics{dens_faz_tog_50k_rez3_SF.eps}}
\resizebox{3.5in}{2.3in} {\includegraphics{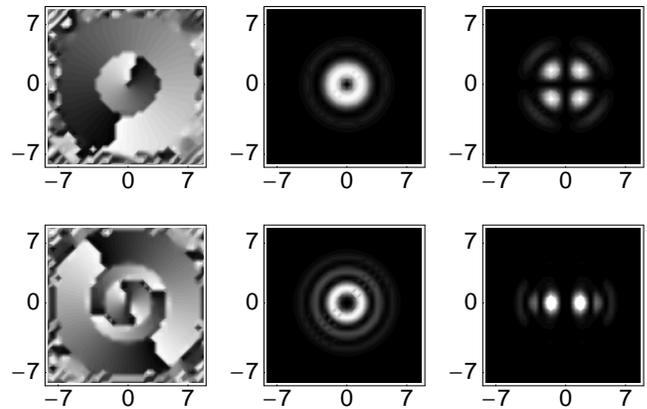}}
\caption[]{Phase (left column) and density (middle and right columns for $xy$ and $xz$ planes, respectively) in $m_F=0$ (upper panel) and $m_F=-1$ (lower panel) components corresponding to the second resonance at maximal population. Initial number of atoms is $N_{+1}=5 \times 10^4$. Other parameters are: $B=-0.16$mG, and $\beta=1$.
\label{ssrad}}
\end{center}
\end{figure}

As it was already mentioned in Section \ref{introduction} the dipolar resonances can be tuned not only by the magnetic field, but also by the trap geometry. In order to demonstrate that, we have performed calculations for the spherically symmetric trap. Two resonances found in this case are shown in Fig. \ref{ss}. The properties of the first resonance (i.e., the one for lower magnetic field) resemble those already discussed for axially symmetric trap. Looking, for example, at the densities in the $xz$ planes we see two and three rings in the $m_F=0$ and $m_F=-1$ components, respectively. However, the situation changes for the second resonance. Here, we observe inner rings structures for both components (see Fig. \ref{ssrad}). This is a signature that also radial excitations not only axial ones start to play a role. The explanation of this is given in Sec. \ref{explanation}.

%fig_6 rezonanse, beta=1

%fig_9 gestosci i faza XY, m_f=0,-1, beta=1, 2-gi rezonans

\section{Origin of dipolar resonances}
\label{explanation}

To understand the origin of dipolar resonances let us first consider collision of two atoms, both initially in $F=1$ and $m_F=+1$ internal state and the ground state of the axially symmetric harmonic oscillator,
\begin{equation}
\varphi_{0} (\boldsymbol{r}) \sim  \mathrm{e}^{-(x^2+y^2 + \beta z^2)/2}
\label{state0}
\end{equation}
with corresponding single particle energy $E_0=1+\beta/2$. Here  $\beta=\omega_z/\omega_{\perp}$ is the aspect ratio of the harmonic trap. The oscillatory units based on the radial frequency $\omega_{\perp}$ are used throughout. We use a basis of single-particle states of harmonic oscillator and assume that contact repulsive interactions do not modify these states substantially. This assumption can be justified if occupation of single particle states is small. Direct evaluation of the two body dipolar matrix elements  leads to the following possible outcomes of the collision:

{\it i) No atom changes its spin projection.} In such a case both atoms stay in the initial state (\ref{state0}). The elastic dipolar collisions contribute effectively to the contact interactions.

{\it ii) Both atoms flip their spin projection by one.} To assure conservation of total angular momentum, the two atom system must gain two quanta of orbital angular momentum. This angular momentum can be shared by both atoms, then each one go to the state with $m_F=0$ and orbital angular momentum $m_L = 1$, or only one atom gains the angular momentum, i.e. it goes to the state with $m_F=0$ and $m_L=2$. Second atom remains in $m_L=0$ state. 

Evidently, in our numerical calculations we do not observe transitions to the state with angular momentum $m_L=2$. We believe that this is due to bosonic nature of atoms which prefers transitions with two atoms going to the same final state. 

{\it iii) Only one atom flips its spin projection.} Conservation of total angular momentum requires that one of the interacting atom has to gain one quanta of angular momentum, i.e. its wave function has $m_L=1$. Due to the symmetry of dipolar interactions the final wave function in $z$ coordinate of relative motion must be antisymmetric. 

For resonances observed in our GP simulations additional rotational quanta is shared by the atom which enters the $m_F=0$ state after the collision. This can be checked by comparing the phases of the $m_F=1$ and $m_F=0$ wave functions. There is no vortex in $m_F=1$ component but there is a single vortex in $m_F=0$ one, see Fig. \ref{denxy} for example.

\begin{figure}
\includegraphics[scale=0.37]{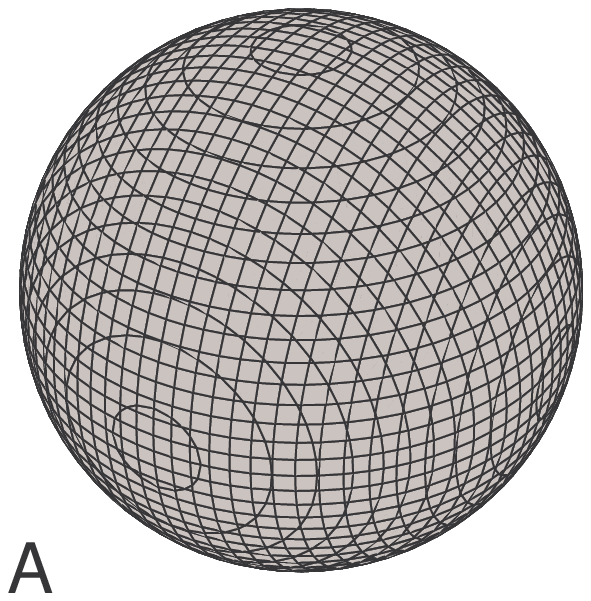}
\includegraphics[scale=0.37]{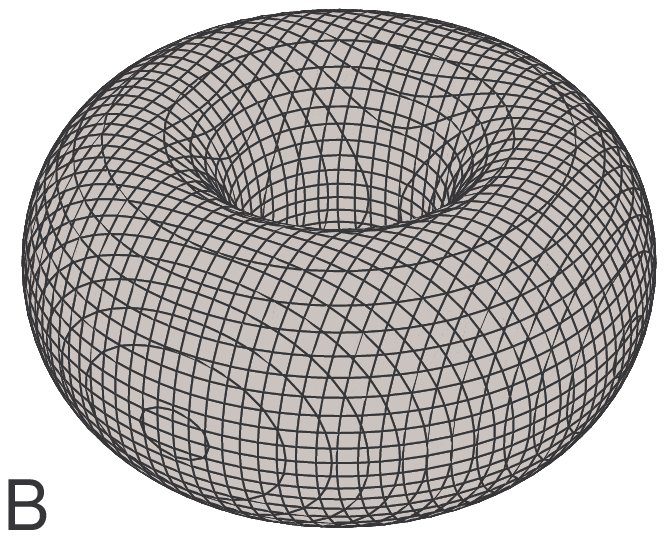}
\includegraphics[scale=0.37]{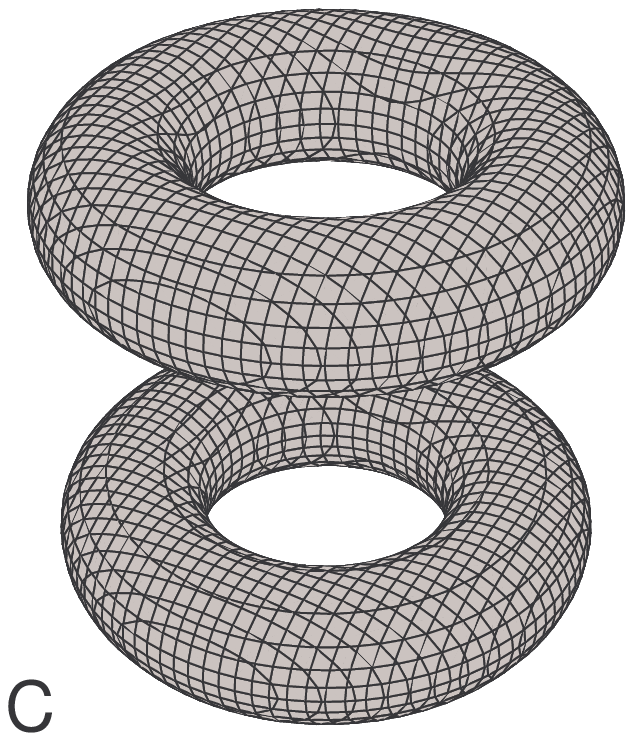}
\caption{Spatial densities of the three lowest states coupled by dipolar interactions. A) ground state of the trap with $m_F=+1$; B) lowest excited state with one quanta of angular momentum and symmetric in the $z$ direction. It is accessible when two interacting atoms flip their spins simultaneously; C) lowest excited state with one quanta of angular momentum and anti-symmetric in the $z$ direction. It is accessible when one ore two interacting atoms flip their spins. } \label{w-functions}
\end{figure}

To gain better understanding of observed density profiles let us introduce a simplistic model where all accessible single-particle states are limited to the lowest energy states of the symmetry consistent with numerical results, i.e. to the following states
\begin{subequations} \label{base3}
\begin{align}
\varphi_{0} (\boldsymbol{r}) &\sim  \mathrm{e}^{-(x^2+y^2 + \beta z^2)/2} \\
\varphi_{1} (\boldsymbol{r}) &\sim (x+iy) \mathrm{e}^{-(x^2+y^2 + \beta z^2)/2} \\
\varphi_{2} (\boldsymbol{r}) &\sim  z(x+iy)\mathrm{e}^{-(x^2+y^2 + \beta z^2)/2} 
\end{align}
\end{subequations}

3D graphs showing characteristic density distributions of all these states are shown in Fig. \ref{w-functions}. Note that $|\varphi_2|^2$ has two density rings located symmetrically with respect to $xy$ plane while $|\varphi_1|^2$ has only one ring in the $z=0$ plane.

Treating dipolar interactions in the first order of the perturbation we can expand the spinor field operator in the basis of these single particle trap states. In a case of the elongated cigar shape trap the lowest excited states are $\varphi_1 (\rho,\phi,z)$ and $\varphi_2 (\rho,\phi,z)$ and we can assume the following form of spinor components of the field operator:
\begin{eqnarray}
\left(
\begin{array}{l}
\hat{\psi}_1 \\
\hat{\psi}_0 \\
\hat{\psi}_{-1}
\end{array}
\right)
=
\left(
\begin{array}{l}
\varphi_0(\boldsymbol{r}) \hat{a}_0\\
\varphi_1(\boldsymbol{r}) \hat{a}_1 + \varphi_2(\boldsymbol{r}) \hat{a}_2\\
0\\
\end{array}
\right)   \;.
\label{spinor-expansion}
\end{eqnarray}
The second quantization Hamiltonian, limited by the expansion Eq.~(\ref{spinor-expansion}) consists of three parts: the single-particle the contact and the dipolar contributions. They can be written as
\begin{subequations}
\begin{equation}
\label{h_dip}
H = H^{SP} + H^C + H^D.
\end{equation}
where the single particle component is:
\begin{align}
H^{SP} &= (E_0+B)\,\Bd\B + E_1\,\Cd\C + E_2\,\Dd\D
\end{align}
and the term proportional to $B$ corresponds to the linear Zeeman shift. The contact two-body term is of the form:
\begin{align}
H^C &= U_{00}\,\Bd\Bd\B\B + U_{11} \,\Cd\Cd\C\C + U_{22}\,\Dd\Dd\D\D \nonumber \\
&+ U_{01}\,\Bd\Cd\B\C+U_{02}\,\Bd\Dd\B\D+ U_{12}\,\Cd\Dd\C\D
\nonumber \\
&+ U_{\mathtt{NR}}\,\Cd\Cd\D\D + U_{\mathtt{NR}}^*\,\Dd\Dd\C\C
\end{align}
and finally the two-body dipolar interaction is:
\begin{align}
H^D &=D_{00}\Bd\Bd\B\B  +  D_{01}\Bd\Cd\B\C + D_{02}\Bd\Dd\B\D \nonumber \\
  &+ T^{(2)}_1(\Cd\Cd\B\B + h.c.)
  + T^{(1)}_2(\Bd\Dd\B\B + h.c.) \nonumber \\
  &+ T^{(2)}_2(\Dd\Dd\B\B + h.c.)
\end{align}
\end{subequations}
All $U$'s, $D$'s, and $T$'s depend on the trap geometry and can be easily evaluated. Exact formulas are given in the Appendix C. Parameters $T^{(P)}_i$ are dipolar integrals for processes when $P=1$ or $2$ particles are transferred from the initial state to the state of wave function $\varphi_i$ ($i=1,2$).

Since total number of particles $N=a_0^{\dagger} a_0 + a_1^{\dagger} a_1 + a_2^{\dagger} a_2$ is conserved, the Hamiltonian separates into diagonal blocks corresponding to a given number of particles.  The ``initial'' state,
\begin{equation}
 |N,0,0\rangle = \frac{1}{\sqrt{N!}}(a_0^{\dagger})^N\, |0\rangle
 \end{equation}
 couples, in the first order perturbation theory, with
\begin{subequations}
\begin{align}
|N-2,2,0\rangle &= \frac{1}{\sqrt{2(N-2)!}}(a_0^{\dagger})^{N-2} (a_1^{\dagger})^2\, |0\rangle \\
|N-1,0,1\rangle &= \frac{1}{\sqrt{(N-1)!}}(a_0^{\dagger})^{N-1} a_2^{\dagger}\, |0\rangle  \\
|N-2,0,2\rangle &= \frac{1}{\sqrt{2(N-2)!}} (a_0^{\dagger})^{N-2} (a_2^{\dagger})^2\, |0\rangle
\label{4states}
\end{align}
\end{subequations}
where $|0\rangle$ is the vacuum.
Let us notice that dipolar matrix element corresponding to two particles going to $\varphi_2({\bf r})$ state is proportional to $\sqrt{2N(N-1)}\,T^{(2)}_2$ which in the limit of large $N$ scales as $\sim N T^{(2)}_2$. The corresponding  matrix element associated with only one particle jumping to this state is $\sqrt{N}(N-1)\,T^{(1)}_2$ and scales with number of particles as $\sim N^{3/2} T^{(1)}_2$. Evidently in the limit of large $N$ atoms enter the  state $\varphi_2({\bf r})$ one by one -- only one atom per every two-body dipolar collision flips its spin. The processes driven by term proportional to $T_i^{(1)}$ dominate the collisions driven by the $T_i^{(2)}$ for large $N$.

For $N=2$ the four states listed above span the whole subspace. The spectrum obtained from the exact diagonalization is presented in Fig. \ref{energy}. The figure explains origin of resonant behavior of the system. For $B=0$ energy of each state is a sum of kinetic energy, potential energy, and contact interaction energy. Energy related to the dipole-dipole interaction is very small and can be ignored. When the magnetic field is switched on the additional Zeeman energy must be taken into account for atoms which are in states with $m_F=+1$. As we see for some magnetic fields the energies of different states become equal and then dipolar interactions start to play a crucial role. First, they cancel degeneracy of the states, second, they can lead to noticeable transfer between them. This is how the dipolar resonances do appear.

Three avoided crossings seen in the Fig. \ref{energy} correspond to three different resonances. The width of every resonance is determined by the magnitude of the dipolar energy and is very small. For two rubidium atoms in a magnetic trap of frequency $\simeq 100$Hz it is of the order of $\Delta B \simeq 10^{-4}$mG. Such a precision in control of the magnetic field is  extremely difficult --  beyond experimental reach.
However, the dipolar energy grows with number of atoms as discussed above. Therefore for $N\sim 10^4$ (as studied in section {\ref{EdH}), widths of resonances increase by several orders of magnitude $\Delta B \simeq 10^{-1}$mG giving a hope for an experimental realization.  Thus for  large systems the effects discussed could be observed in a rubidium condensate in experiments performed at ultra-low magnetic fields requiring a special shielding of all external magnetic fields. On the other hand, large atomic densities can be reached in optical lattices with few atoms per site only. In such a situation chromium atoms are preferred because of relatively large magnetic moment. For typical optical lattices the width of resonances for two chromium atoms in a single site is about $\Delta B \simeq 10^{-1}$mG. Moreover, in the case of optical latice, the width of the resonance is significantly increased due to the finite energy width of excited bands as seen in experiment \cite{Bruno}. Again special shielding is required for experimental observation of resonant dipolar interactions. If the magnetic field does not match a resonant value, the dipolar interactions can be totally ignored.

Weakness of dipolar interactions is responsible for a very small width of the resonances making them difficult to detect. On the other hand weak dipolar interactions have one very important advantage: all the resonances shown in Fig. \ref{energy} are well separated. Choosing an appropriate value of the magnetic field a particular resonance can be addressed, and dipolar interactions can be tuned to couple desired center of mass states. 
Obviously far from all the resonances the dipolar interaction can be set to zero. 

Exact value of the resonant magnetic field depends on energy difference of resonantly coupled states at $B=0$ and it is a linear function of number of atoms in the initial state. For realistic values of scattering lengths the contact interaction in the initial $|N,0,0\rangle$ state grows with $N$ faster then the contact interaction in three other considered here states. Therefore, the total energy of this state can be larger then the energy of the resonantly coupled state.  It clearly follows from Fig. \ref{energy} that for large $N$ the corresponding resonant magnetic field is  negative, i.e. its direction is opposite to the direction of the magnetic moment of atoms. This simple observation explains why in our numerical calculations discussed in sec. \ref{EdH} a direction of the magnetic field was inverted with respect to the direction used for the preparation of the ground state.

\begin{figure}[!htb]
\begin{center}
\resizebox{3.1in}{2.1in} {\includegraphics{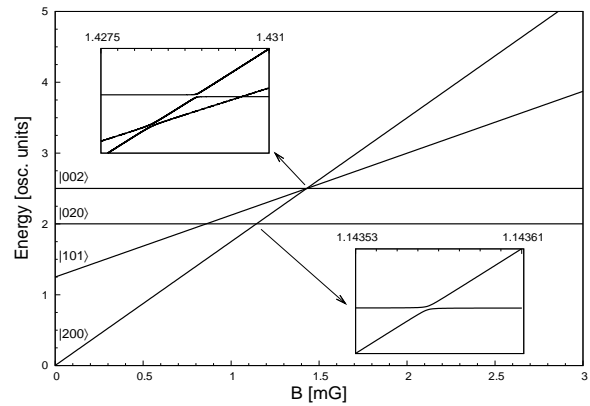}}
\caption[]{Energy of two-atom quantum states as a function of external magnetic field for $\omega_z/\omega_\perp = 1/4$. For some values of external magnetic field energies of some states become equal due to the linear Zeeman effect. Two insets show real and avoiding crossings. Note that avoiding crossings appear only for states coupled by dipolar interactions. 

\label{energy}}
\end{center}
\end{figure}

\section{Conclusions}
\label{conclusions}

In conclusion, we have shown the existence of dipolar resonances in rubidium spinor condensates. The resonances occur when the Zeeman energy of atoms in $m_F=1$ component matches the excitation energy of a spatial state which is populated upon the transfer. Symmetries of dipolar interactions force the atoms in $m_F=0,-1$ states to circulate around the quantization axis. In fact, singly and doubly quantized vortices are formed in $m_F=0,-1$ components, respectively. Since dipolar interactions are weak the resonances are narrow. It means that dipolar resonances are very selective, i.e. one can choose the transition to the proper state by choosing the magnetic field. Hence, dipolar resonances seem to be a route to the observation of the Einstein-de Haas effect, as well as other phenomena related to dipolar interactions, in weak dipolar systems.

\acknowledgments
We are grateful to K. Bongs and K. Rz\c a\.zewski for helpful discussions. J.Z. i M.G. acknowledge hospitality from ICFO and partial support from the advanced ERC-grant QUAGATUA. This work was also supported by the UE project NAME-QUAM and Polish Ministry for Science and Education for 2009-2011 and via grant N202 124736 for 2009-2012 (J.Z.). M.L. acknowledges also  ERC Grant QUAGATUA, Spanish MICINN Projects (FIS2008-00784 and QOIT), EU Grants (AQUTE, NAMEQUAM), and Humboldt Foundation.

\appendix
\section{Equation of motion}
\label{first}

The equation of motion (\ref{Eqmot}) is obtained from the Heisenberg equations for the field operators $\hat{\psi}_i(\boldsymbol{r})$ for the system described by the Hamiltonian, $H$ (\ref{Ham}). Since the Hamiltonian (\ref{Ham}) splits in single-particle part and two-particle parts related to contact and dipolar interactions, the commutators $[\hat{\psi}_i(\boldsymbol{r}),H]$ as well as the right hand-site of Eq. (\ref{Eqmot}) get the form of the sum of three terms of different physical origin. The single-particle part is diagonal and given by ${\cal{H}}^{sp}=-\frac{\hbar^2}{2 m} \nabla^2 + V_{ext} - \mu m_F B$.

The term ${\cal{H}}^c$ in Eq. (\ref{Eqmot}), which is related to the contact interactions has the following diagonal elements
\begin{eqnarray}
{\cal{H}}^c_{11} &=& (c_0+c_2)\, \hat{\psi}^{\dagger}_1 \hat{\psi}_1
+(c_0+c_2)\, \hat{\psi}^{\dagger}_0 \hat{\psi}_0   \nonumber \\
&+&  (c_0-c_2)\, \hat{\psi}^{\dagger}_{-1} \hat{\psi}_{-1}  \nonumber  \\
{\cal{H}}^c_{00} &=&(c_0+c_2)\, \hat{\psi}^{\dagger}_1\hat{\psi_1} +
c_0\, \hat{\psi}^{\dagger}_0\hat{\psi_0}  \nonumber \\
&+& (c_0+c_2)\, \hat{\psi}^{\dagger}_{-1}\hat{\psi_{-1}} \nonumber \\
{\cal{H}}^c_{-1-1} &=&(c_0-c_2)\, \hat{\psi}^{\dagger}_1\hat{\psi}_1
+(c_0+c_2)\, \hat{\psi}^{\dagger}_0\hat{\psi}_0  \nonumber \\
&+& (c_0+c_2)\, \hat{\psi}^{\dagger}_{-1} \hat{\psi}_{-1}  \;.
\label{Hc}
\end{eqnarray}
These elements describe the collisions of atoms that preserve the projection of the spin for each atom. The off-diagonal elements, on the other hand, are responsible for collisions changing separately the atomic spin projections but conserving the projection of total spin. They yield
\begin{eqnarray}
{\cal{H}}^c_{10} &=& c_2\hat{\psi}^{\dagger}_{-1} \hat{\psi}_0  \nonumber \\
{\cal{H}}^c_{0-1} &=& c_2\hat{\psi}^{\dagger}_0\hat{\psi}_1  \nonumber \\
{\cal{H}}^c_{1-1} &=& 0  \;.
\end{eqnarray}

For dipolar interactions, ${\cal{H}}^d$, in Eq. (\ref{Eqmot}) one has
\begin{eqnarray}
{\cal{H}}^d_{ij}(\boldsymbol{r}) &=& \int d^3r' \hat{\psi}_k^{\dagger}(\boldsymbol{r}')
V^d_{ij,kl}(\boldsymbol{r}-\boldsymbol{r}')   \hat{\psi}_l(\boldsymbol{r}')   \;,
\label{Hd}
\end{eqnarray}
where
\begin{eqnarray}
&&V^d_{ij,kl}(\boldsymbol{r}-\boldsymbol{r}') = \frac{\mu^2}{|\boldsymbol{r}-\boldsymbol{r}'|^3}
\boldsymbol{F}_{ij} \boldsymbol{F}_{kl}  \nonumber \\
&&- \frac{3\mu^2}{|\boldsymbol{r}-\boldsymbol{r}'|^5}
[\boldsymbol{F}_{ij}  (\boldsymbol{r}-\boldsymbol{r}')] [\boldsymbol{F}_{kl}  (\boldsymbol{r}-\boldsymbol{r}')]
\,\,.
\label{Vd}
\end{eqnarray}
Only two elements of ${\cal{H}}^d$ matrix are independent, namely
\begin{eqnarray}
&&{\cal{H}}^d_{11}(\boldsymbol{r}) = \mu^2 \int d^3r' \left[
\frac{1}{|\boldsymbol{r}-\boldsymbol{r}'|^3}-3\frac{(z-z')^2}{|\boldsymbol{r}-\boldsymbol{r}'|^5}
\right] \nonumber \\
&&\times (\hat{\psi}_1^{\dagger} \hat{\psi}_1-\hat{\psi}_{-1}^{\dagger} \hat{\psi}_{-1})
\nonumber \\
&& -3 \frac{\mu^2}{\sqrt{2}}
\int d^3r' \frac{z-z'}{|\boldsymbol{r}-\boldsymbol{r}'|^5}[(x-x') - i (y-y')]  \nonumber \\
&&\times (\hat{\psi}_1^{\dagger}\hat{\psi}_0+\hat{\psi}_0^{\dagger}\hat{\psi}_{-1})    \nonumber \\
&& -3 \frac{\mu^2}{\sqrt{2}}
\int d^3r' \frac{z-z'}{|\boldsymbol{r}-\boldsymbol{r}'|^5}[(x-x') + i (y-y')] \nonumber \\
&&\times (\hat{\psi}_0^{\dagger}\hat{\psi}_1+\hat{\psi}_{-1}^{\dagger}\hat{\psi}_0)
\label{Hd_11}
\end{eqnarray}
and
\begin{eqnarray}
&&{\cal{H}}^d_{10}(\boldsymbol{r}) = -3 \frac{\mu^2}{\sqrt{2}} \int d^3r'
\frac{[(x-x')-i(y-y')](z-z')}{|\boldsymbol{r}-\boldsymbol{r}'|^5} \nonumber \\
&& \times (\hat{\psi}_1^{\dagger} \hat{\psi}_1-\hat{\psi}_{-1}^{\dagger} \hat{\psi}_{-1})
\nonumber \\
&& -\frac{3}{2} \mu^2 \int d^3r'
\frac{[(x-x')-i(y-y')]^2}{|\boldsymbol{r}-\boldsymbol{r}'|^5}
(\hat{\psi}_1^{\dagger}\hat{\psi_0}+\hat{\psi_0}^{\dagger}\hat{\psi}_{-1})   \nonumber \\
&& +\mu^2 \int d^3r' \left[
\frac{1}{|\boldsymbol{r}-\boldsymbol{r}'|^3}-\frac{3}{2}\frac{(x-x')^2+(y-y')^2}
{|\boldsymbol{r}-\boldsymbol{r}'|^5} \right]   \nonumber \\
&& \times (\hat{\psi}_0^{\dagger}\hat{\psi}_1+\hat{\psi}_{-1}^{\dagger}\hat{\psi}_0)   \;.
\label{Hd_10}
\end{eqnarray}
For other elements one has
\begin{eqnarray}
&&{\cal{H}}^d_{0-1} = {\cal{H}}^d_{10}  , \;\;\;
{\cal{H}}^d_{-1-1} = - {\cal{H}}^d_{11}   \nonumber \\
&& {\cal{H}}^d_{1-1} = {\cal{H}}^d_{00} = 0  \,.
\label{Hd_con}
\end{eqnarray}

The ${\cal{H}}^d_{ij}$ terms (both diagonal and off-diagonal) are responsible for the change of the total spin projection of colliding atoms. For example, even if components $m_F=0$ and $m_F=-1$ are initially empty then due to the first part in ${\cal{H}}^d_{10}$ there exists a process when one of two atoms being initially in $m_F=1$ state populates $m_F=0$ state after collision.

\section{Numerical details}
\label{second}

To solve Eq. (\ref{Eqmot}) we neglect the quantum fluctuations and replace
the field operator $\hat{\psi}_i(\boldsymbol{r})$ ($\hat{\psi}_i^{\dagger}(\boldsymbol{r})$)
by an order parameter $\psi_i(\boldsymbol{r})$ ($\psi_i^*(\boldsymbol{r})$) for each component.
Next, we split the right-hand side operator of (\ref{Eqmot}) in the following way
\begin{equation}
{\cal{H}}={\cal{H}}^{sp} + {\cal{H}}^c + {\cal{H}}^d = T + V   \;,
\end{equation}
where $T$ is the kinetic energy operator which is a diagonal matrix and $V$ represents the
potential energy including the trapping potential, the interaction with the external
magnetic field, and the contact and dipolar interactions. Now, we apply the split-operator
method to solve the mean-field version of Eq. (\ref{Eqmot}).
First, we calculate the evolution of an order parameter
according to the potential energy. To this end, we diagonalize the
matrix $V$ at each spatial point $V=P D P^{-1}$, where the matrix $P$
consists of the eigenvectors  written in columns and the matrix $D$ is
diagonal with the eigenvalues on the diagonal. The evolution by the
time step $\Delta t$ is given by
\begin{eqnarray}
e^{-i V \Delta t /\hbar}
\left(
\begin{array}{l}
{\psi}_1 \\
{\psi}_0 \\
{\psi}_{-1}
\end{array}
\right)
= P e^{-i D \Delta t /\hbar} P^{-1}
\left(
\begin{array}{l}
{\psi}_1 \\
{\psi}_0 \\
{\psi}_{-1}
\end{array}
\right)
\label{evo}
\end{eqnarray}
and is calculated in three steps: first, the spinor order parameter is multiplied by the matrix $P^{-1}$, next by $e^{-i D \Delta t /\hbar}$, which is a diagonal matrix since $D$ is diagonal, and finally by the matrix $P$. Then we calculate the evolution due to the kinetic part. Since the kinetic energy operator $T$ is diagonal we apply the Fourier technique to each component in a way as it is usually done in the case of a scalar Gross-Pitaevskii equation.

All integrals appearing in formulas (\ref{Hd_11}) and (\ref{Hd_10}) are the convolutions and we use the Fourier transform technique to calculate them. The Fourier transform of the convolution of two functions is the product of the Fourier transforms of these functions. Therefore, it is useful to have analytical formulas for the Fourier transforms of the components of the convolutions that do not change during the condensate evolution. The appropriate Fourier transforms are found in the following way. First, two rotations of coordinates system (by angles: $\beta$ around $z$-axis and $\alpha$ around $y$-axis) are applied to make the vector $\boldsymbol{k}$ -- the argument of the Fourier transform -- parallel to the $z$-axis and next, the regularization procedure as described in Ref. \cite{Goral} is used. The result reads
\begin{eqnarray}
&&{\cal{F}} \left[
\frac{1}{|\boldsymbol{r}|^3}-3\frac{z^2}{|\boldsymbol{r}|^5} \right]
= -\frac{4\pi}{3}(1-3 \cos^2\!\alpha)   \nonumber  \\
&&{\cal{F}} \left[
\frac{1}{|\boldsymbol{r}|^3}-\frac{3}{2}\frac{x^2+y^2}{|\boldsymbol{r}|^5} \right]
= \frac{2\pi}{3}(1-3 \cos^2\!\alpha)   \nonumber  \\
&&{\cal{F}} \left[
\frac{(x-iy) z}{|\boldsymbol{r}|^5} \right] =
\frac{2\pi}{3}\, e^{-i \beta} \sin 2 \alpha   \nonumber  \\
&&{\cal{F}} \left[
\frac{(x+iy) z}{|\boldsymbol{r}|^5} \right] =
\frac{2\pi}{3}\, e^{i \beta} \sin 2 \alpha   \nonumber  \\
&&{\cal{F}} \left[
\frac{(x-iy)^2}{|\boldsymbol{r}|^5} \right] = -\frac{4\pi}{3}\,  e^{-i 2\beta}
\sin^2\!\alpha    \,\,,
\label{FT}
\end{eqnarray}
with the angles $\alpha$ and $\beta$ defined as follows:
$\cos \alpha = k_z /k$ and $\sin \beta = k_y /\sqrt{k^2 - k_z^2}$.

Even with a help of formulas (\ref{FT}) the evolution of mean-field version of Eq. (\ref{Eqmot}) is a time-consuming computational task. We would like to point out that each time step consists of taking nine numerical Fast Fourier Transforms (three for determining the evolution of the spinor wave function governed by the kinetic energy and six for calculating ${\cal{H}}^d$ matrix) and also doing the diagonalization of matrix $V$ at each spatial point. Therefore the parallel computing with OpenMP directives has been used in our calculations.

Size of the grid was adjusted to the trap geometry -- we have chosen $2^5\times 2^5\times 2^5$ grid points in the case of spherically symmetric traps (with the spatial steps equal to $\Delta x= \Delta y= \Delta z= 0.55$ osc. units, where osc. unit=$\sqrt{\hbar/m\omega}$) and $2^5\times 2^5\times 2^6$ grid points for cigar-like traps (with the spatial steps equal to $\Delta x=\Delta y=0.275$ and $\Delta z=0.375$ osc. units). The time step was chosen to $\Delta t=2.5\times 10^{-5}$ osc. units, where osc. unit=$1/\omega$. Part of the results was recomputed with twice smaller spatial step (two times bigger grid size in every dimension) as a self-test.

The ground state was obtained by evolving the mean-field version of Eq. (\ref{Eqmot}) in imaginary time at the value of magnetic field $B=0.73$ mG. Such magnetic field was large enough to result in having almost all atoms in $m_F=1$ Zeeman sublevel.

\section{Two-body matrix elements}

In this section we present explicit formulas for two-body matrix elements of the Hamiltonian \eqref{Ham} in the basis defined by \eqref{base3}.

Contact integrals:
\begin{subequations}
\begin{align}
U_{00} &= \frac{c_0+c_2}{2}\int\!\!\mathrm{d}^3r\,\,|\varphi_0(\boldsymbol{r})|^4 \\
U_{11} &= \frac{c_0}{2}\int\!\!\mathrm{d}^3r\,\,|\varphi_1(\boldsymbol{r})|^4 \\
U_{22} &= \frac{c_0}{2}\int\!\!\mathrm{d}^3r\,\,|\varphi_2(\boldsymbol{r})|^4 \\
U_{01} &= (c_0+c_2)\int\!\!\mathrm{d}^3r\,\,|\varphi_0(\boldsymbol{r})|^2|\varphi_1(\boldsymbol{r})|^2 \\
U_{02} &= (c_0+c_2)\int\!\!\mathrm{d}^3r\,\,|\varphi_0(\boldsymbol{r})|^2|\varphi_2(\boldsymbol{r})|^2 \\
U_{12} &= 2c_0\int\!\!\mathrm{d}^3r\,\,|\varphi_1(\boldsymbol{r})|^2|\varphi_2(\boldsymbol{r})|^2 \\
U_{\mathtt{NR}} & =
\frac{c_0}{2}\int\!\!\mathrm{d}^3r\,\,\Big[\varphi^*_1(\boldsymbol{r})\varphi_2(\boldsymbol{r})\Big]^2
\end{align}
\end{subequations}

Dipolar interactions which do not change the spin of atoms:
\begin{subequations}
\begin{align}
D_{00} &= \frac{\mu^2}{2}\int\!\!\int\!\!\mathrm{d}^3r\,\mathrm{d}^3r'\,\, |\varphi_0(\boldsymbol{r})|^2 f_1(\boldsymbol{r}-\boldsymbol{r'})|\varphi_0(\boldsymbol{r'})|^2 \\
D_{01} &= -\frac{\mu^2}{2}\int\!\!\int\!\!\mathrm{d}^3r\,\mathrm{d}^3r'\,\, \varphi_1^*(\boldsymbol{r})\varphi_0(\boldsymbol{r}) f_1(\boldsymbol{r}-\boldsymbol{r'})\varphi_1^*(\boldsymbol{r'})\varphi_0(\boldsymbol{r'})\\
D_{02} &= -\frac{\mu^2}{2}\int\!\!\int\!\!\mathrm{d}^3r\,\mathrm{d}^3r'\,\, \varphi_2^*(\boldsymbol{r})\varphi_0(\boldsymbol{r}) f_1(\boldsymbol{r}-\boldsymbol{r'})\varphi_2^*(\boldsymbol{r'})\varphi_0(\boldsymbol{r'})
\end{align}
\end{subequations}

Dipolar interactions which change the spin of the atoms:
\begin{subequations}
\begin{align}
T_1^{(2)} &= \frac{\mu^2}{2}\int\!\!\int\!\!\mathrm{d}^3r\,\mathrm{d}^3r'\,\, \varphi_1^*(\boldsymbol{r})\varphi_0(\boldsymbol{r}) f_2(\boldsymbol{r}-\boldsymbol{r'})\varphi_1^*(\boldsymbol{r'})\varphi_0(\boldsymbol{r'})\\
T_{2}^{(1)} &= \mu^2\int\!\!\int\!\!\mathrm{d}^3r\,\mathrm{d}^3r'\,\, \varphi_0^*(\boldsymbol{r})\varphi_0(\boldsymbol{r}) f_3(\boldsymbol{r}-\boldsymbol{r'})\varphi_2^*(\boldsymbol{r'})\varphi_0(\boldsymbol{r'}) \\
T_{2}^{(2)} &= \frac{\mu^2}{2}\int\!\!\int\!\!\mathrm{d}^3r\,\mathrm{d}^3r'\,\, \varphi_2^*(\boldsymbol{r})\varphi_0(\boldsymbol{r}) f_2(\boldsymbol{r}-\boldsymbol{r'})\varphi_2^*(\boldsymbol{r'})\varphi_0(\boldsymbol{r'})
\end{align}
\end{subequations}

Functions $f_i$ have a form:
\begin{subequations}
\begin{align}
f_1(\boldsymbol{r}) &= \frac{1-3 n_z^2}{r^3} \\
f_2(\boldsymbol{r}) &= -\frac{3}{2}\frac{(n_x-in_y)^2}{r^3} \\
f_3(\boldsymbol{r}) &= -\frac{3\sqrt{2}}{2}\frac{n_z(n_x+in_y)}{r^3}
\end{align}
\end{subequations}

\end{document}